\documentclass[aps,preprint,amssymb]{revtex4}
\usepackage[hypertex,backref]{hyperref}
\usepackage{epsfig}
\usepackage{amsmath,amsfonts,amssymb}
\usepackage{xcolor}

\newcommand{\f}{\frac}

\newcommand{\m}{\mathbf}

\newcommand{\eqlref}[1]{Eq.~\eqref{#1}}

\begin{document}
\title{The nature of the rectilinear diameter singularity}
\author{V.L. Kulinskii$^a$,  N. P. Malomuzh$^a$}
\email[]{E-mail: kulinskij@onu.edu.ua}%
\affiliation{Department for Theoretical Physics,
Odessa National University, Dvoryanskaya 2, 65026 Odessa, Ukraine}
\date{\today}
\begin{abstract}
The rigorous explanation for the term $\vert t \vert^{2\beta}$ in
the rectilinear diameter equation is given ($t = (T_c-T)/T_c$,
$\beta$ is the critical exponent for the asymptotic form of the
equation of state). The optimal order parameter, for which the
branches of binodal are symmetric is constructed within the
canonical formalism. It is shown that the ratio of the amplitudes
for the diameter singularity of the order parameter
$\f{D_{1-\alpha}}{D_{2\beta}}$ before $|t|^{1-\alpha}$ and $|t|^{2\beta}$, where $\alpha$
determines the behavior of the heat
capacity and $\beta$ is the critical exponent of the order parameter,
takes the universal character modulo nonuniversal factor
which depends on the thermodynamic class of the corresponding
states. The analysis of entropy for argon and water leads to $\beta
= 0.33$ and the corresponding amplitude ratio $\f{S_{1-\alpha}}{S_{2\beta}}\approx -3.5$.
\end{abstract}
\maketitle
\section{Introduction}
The nature of the asymmetry of the coexistence curve in liquids
and multi-component solutions has a long history, which dates back
to the work of Cailletet and Matthias \cite{diam0}, where the
empirical linear law was obtained (see also
\cite{diamexp/molphys/1986} and references therein). The diameter
$\varphi ^{(d)}(t)$ is determined as following:
\begin{equation}
\label{eq1} \varphi ^{(d)}(t) = \frac{1}{2}(\varphi ^{(l)}(t) +
\varphi ^{(g)}(t)),
\end{equation}
where $\varphi ^{(i)}(t)$, $i = l,g,$ are the values of the order
parameter on the liquid and gas branches of the coexistence curve,
considered as a function of the dimensionless temperature $t =
{\frac{{T - T_{c}}} {{T_{c}}} }$ ($T_{c} $ is the critical
temperature). According to the fluctuation theory of critical
phenomena \cite{patpokr} $\varphi ^{(d)}(t)$ is described by the
expansions:
\begin{equation}
\label{eq2} \varphi ^{(d)}(t) = {\left\{ {{\begin{array}{*{20}c}
 {D_{1} t + D_{2}\,t^{2} + \ldots\,\,,} \hfill & {\vert t \vert
 \gg
t_{G}}  \hfill \\
 {D_{1-\alpha} \vert t\vert ^{1 - \alpha}  + \ldots,}
 \hfill & {\vert t \vert \ll t_{G}}  \hfill \\
\end{array}}}  \right.},
\end{equation}
where $t_{G} $ is the Ginzburg temperature, and $\alpha $ is the
critical exponent for the heat capacity. The ``$\vert
t\vert^{1-\alpha}$`` anomaly, where $t=\f{T-T_c}{T_c}$, of the
rectilinear diameter was predicted by Widom and Rowlison in
\cite{diamrowidom/jcp/1970}. Since $1 - \alpha $ is close to unit
($ \approx 0.89)$, the influence of the singular term in
(\ref{eq2}) is expected to be essential only in the intermediate
vicinity of the critical point. However, the careful analysis of
experimental data for the rectilinear diameter may display another
type of the critical singularity:
\begin{equation}
\label{eq3} \varphi ^{(d)}(\tau )= D_{2\beta} \vert \tau \vert
^{2\beta} + D_{1-\alpha} \vert \tau \vert ^{1 - \alpha}  + D_{1}
\tau + D_{2} \tau ^{2} + \ldots,
\end{equation}
where $\beta$ is the critical exponent for coexistence curve:
$2\beta \approx 0.654 < 1 - \alpha $ (for the review see
\cite{diamexp/molphys/1986}). Because of very close values of the
exponents of the first three terms one needs the data of high
precision to get the reliable estimates for the amplitudes $D_i$.
Here we use the temperature like variable $\tau$, which depends on
the reduced temperature $t$ and will be specified below.

The rectilinear diamenter singularity has intensively discussed in
\cite{yydiamfisherorkoulas/prl/2000,fishmixdiam1/pre/2003} in
connection with the nature of ``$\vert t\vert^{2\beta}$``\,
anomaly and its relation with the Yang-Yang (YY) singularity
\cite{yang2/prl/1964}. As was shown in \cite{rehrmermin/pra/1973}
the appearance of the $|t|^{2\beta}$ term in the rectilinear
diameter demands that the equation of state is as following:
\begin{equation}\label{pmutinv}
  \pi(\zeta,\tau) = \pi_0 + \tau^{2-\alpha}\,
  f_{\pm} \left(\,\f{\zeta }{\tau^{\beta +\gamma}} \,\right)\,,\quad
  \tau  = \tau (P,T,\mu)\,,\quad \zeta  = \zeta (P,T,\mu)\,,
\end{equation}
where $\pi$, $\tau$ and $\zeta $ are the analytic functions of all
three thermodynamic fields $P,\mu,T$. Here $\zeta$ is the field
conjugated to the order parameter, which in general is not
density, $\tau$ is the ``temperature``-like variable and $\pi$ is
the thermodynamic potential corresponding to them (the details see
in \cite{rehrmermin/pra/1973}). Thus the scaling law
\eqref{pmutinv} states that there is no preferred variable between
$P$, $\mu$ and $T$ \cite{griffitswheeler/pra/1970}. Within the
approach proposed in
\cite{yydiamfisherorkoulas/prl/2000,fishmixdiam1/pre/2003} the YY
anomaly as well as ``$\vert t\vert^{2\beta}$``\, singularity
appear as the result of the nonlinear dependence of the physical
quantities on the scaling fields due to \textit{complete/pressure mixing},
which is postulated. However, the regular method to construct the
variables $\zeta$ and $\tau$ was not proposed.

Here we will not discuss the appearance of the $|t|^{2\beta}$ due to
nonlinear transformation of the laboratory order parameter (of type
$v = 1/n$, where $n$ is the density and $v$ is the specific volume).
It is clear that e.g. for the Ising model one is free to chose the ``non symmetrical`` order parameter $\tilde{m} = m+\gamma_2 m^2$, where $m$ is the magnetization,  which gives the corresponding $\tau^{2\beta}$ term for the rectilinear diameter.
The
choice of the proper order parameter in binary mixtures which
restores the symmetry of the binodal was discussed in
\cite{schroer/jchemp/1999,schroer/pccp/2003}, though any procedure
for constructing such an order parameter was not given.
The problem is to find such a microscopic observable $\eta$ so that:
\[
\left\langle\, \eta \,\right\rangle_{1} = -\left\langle\, \eta \,\right\rangle_{2}
\]
along the coexistence curve.
Then in the framework of the RG theory the asymptotic equation of state
is fully symmetrical:
\begin{equation}\label{eoscal}
  \left\langle\,\eta\,\right\rangle =
  \pm \,\tau^{\beta}\,g_{s} \left(\,\f{h}{\tau^{\beta + \gamma }} \,\right)
\end{equation}
where $h$ and $\tau$ are the external field and temperature like
scaling fields correspondingly. However the direct comparison of
the theoretical prediction with the experimental data is
impossible since the explicit form of the order parameter $\eta$
and the fields $h$ and $\tau$ remain to be unstated.

The symmetrical form of the Eq.~\eqref{eoscal} is based on the
symmetry property of the effective Hamiltoniian
$\mathcal{H}[\varphi(\mathbf{r})] =
\mathcal{H}[-\varphi(\mathbf{r})]$ along the binodal. In particular the Landau-Ginzburg Hamiltonian satisfies this condition \cite{patpokr}.

In this paper we give the rigorous formalism for the construction
of the scaling fields $\tau$ and $h$. Within such a formalism the
nonlinear mixing of the laboratory fields arises naturally. In
particular we show that both $|t|^{2\beta}$ and $|t|^{1-\alpha}$
anomalies are of the same nature.

Further we demonstrate this statement using the entropy as the order
parameter. It is natural that the entropy is more sensitive to the
asymmetry in the ``particle-hole`` configuration  than the density.
The latter is just the
average of 1-particle distribution function for the particles.

The canonical transformation $\varphi \to \eta(\varphi) $ \,of the
initial (laboratory) order parameter $\varphi $ is the central
point of the proposed approach. Due to this the local part of the
fluctuation Hamiltonian of liquids reduces to the canonical form,
which is identical to the local part of the Landau-Ginzburg (LG)
Hamiltonian. In canonical variables the coexistence curve is fully
symmetrical, as it takes place for the Ising model. The non-zero
values of the rectilinear diameter arise as the result of the
inverse transformation $\eta \to \varphi $ to the laboratory order
parameter. In the framework of the canonical formalism the
isomorphism of critical fluctuations in liquids and the Ising
model is manifested in the similarity of their thermodynamic
potentials as well as their Hamiltonians. Within the perturbative
approach it can be shown that the asymmetry in main order depends
on the value of the coefficient $a_{5} $ at $\varphi ^{5}$ term in
the initial Hamiltonian. This is accordance with the results of
\cite{nicoll/pra/1981}.


The construction of the canonical order parameter $\eta$ for which
$\eta ^{(d)} = 0$ is discussed below.

\section{Canonical formalism}
The
Hamiltonian for the real fluids can be represented as following:
\begin{equation}\label{hreal}
H[\varphi(\mathbf{r})]=\int \left(\,h_{l}(\varphi(\mathbf{r})) + h_{ql}
\left(\varphi(\mathbf{r})\right)\,\right) dV\,\,.
\end{equation}
where
\[h_{l}(\varphi(\mathbf{r}))=
\sum\limits_{n=1}^{\infty}\frac{a_{n}}{n} \,
\varphi^{n}(\mathbf{r})\] is the local density and
\[h_{ql}(\varphi(\mathbf{r}))=
\sum\limits_{n=1}^{\infty}\frac{a_{n,2}}{n} \,
\varphi^{n}\,\left(\,\nabla \varphi\,\right)^2 \,\]
is the quasilocal
part in small gradient approximation.
The Hamiltonian \eqref{hreal} includes odd power terms of the
order parameter responsible for the ``asymmetry effect`` in the
critical behavior of the system. It should be noted that there are two kinds of
asymmetry terms in the Hamiltonian. They are the local and the quasilocal terms of odd powers.
The importance of the lasts for the RDS was shown in
\cite{nicoll/pra/1981,nicollzia/prb/1981}.

Note that the local term $\varphi^5$ leads to the mixing of the
initial thermodynamic variables $\mu$ and $T$. The quasilocal term
$\varphi(\nabla \varphi)^2$ leads to the $1-\alpha$ singularity of
the rectilinear diameter, though $2\beta$ anomaly was not obtained
there.

Now let us appeal to the ideology of the Catastrophe
Theory. It gives the ground to reduce the initial Hamiltonian given by the infinite series \eqlref{hreal} to the canonical form. For this purpose the order parameter should be subjected to the so called local canonical transformation
\begin{equation}\label{etaphi}
  \eta(\m{r}) = \varphi(\m{r}) + \f{1}{2}\,\Gamma_2\,\varphi^2(\m{r})+\f{1}{3}\,\Gamma_3\,\varphi^3(\m{r}) \ldots\,\,.
\end{equation}
%
As a result the local density of the Hamiltonian takes the form:
\begin{equation}\label{hcan}
h^{(can)}_{l}(\eta)=A_{1}\eta(\mathbf{r})+\frac{A_{2}}{2}\,\eta^{2}(\mathbf{r})+
\frac{A_{4}}{4}\,\eta^{4}(\mathbf{r})\,.
\end{equation}
with $A_4 = const$.
Using the procedure described in \cite{canformkulinskii/jmolliq/2003} one can show that
\begin{align}\label{gamma}
  \Gamma_2 =&\, A_1-a_1\,,\notag\\
  \Gamma_3 =&\,
\f{1}{2}\,A_2-\f{1}{2}\,A^2_1+\f{3}{2}\,a_1\,\Gamma_2-\f{1}{2}\,a_2+\f{1}{2}\,a_1^{2}+\f{3}{2}\,\Gamma^2_2
\,,
\end{align}
\begin{multline}
\Gamma_4 =-\f{1}{2}\,A_1\,A_2+\f{1}{6}\,A^3_1-\f{5}{2}\,a_1\,\Gamma^2_{2}\\ + a_2\,\Gamma_2-a^2_1 \Gamma_2+\f{4}{3}\,a_1\,\Gamma_3-\f{1}{3}\,a_3+\f{1}{2}\,a1\, a_2\\-\f{1}{6}\,a^3_1-\f{5}{2}\,\Gamma^3_2+\f{10}{3}\,\Gamma_3\,\Gamma_2
\,,
\end{multline}
In this paper we illustrate the explicit construction of the transformation \eqref{etaphi} which reduce only the local part of the Hamiltonian to the canonical form.
In such a case the
coefficients $A_{1} $ and $A_{2} $ should vanish simultaneously with
$a_{1} $ and $a_{2}$:
\begin{equation}\label{a1a2}
A_{1} = 0\,, \quad A_{2} = 0 \Leftrightarrow a_1= 0\,,\quad a_2 = 0\,.
\end{equation}
These conditions correspond to the invariance of the locus of the CP in mean-field approximation.

To find the explicit expressions for the coefficients $A_{1} ,A_{2}
$ and $A_{4} $ we demand the fulfillment of the following conditions:
a) the conservation of the local partition function:
\begin{equation}
\label{eqpartition} \int\limits_{ - \infty} ^{\infty}  dx\exp \left(
- h^{(can)}_{loc}(x)\right)  = \int\limits_{ - \infty} ^{\infty} dy\exp
\left( - h_{loc} (y)\right)\,.
\end{equation}
b) one-to-one correspondence (bijectivity) of the canonical transformation \eqref{etaphi}:
\begin{equation}\label{cantransint}
\int\limits^{\eta}_{-\infty}\exp\left(-h^{(can)}_{loc}(z)\right)dz
=\int\limits^{\varphi}_{-\infty}\exp\left(-h_{l}(z)\right)dz\,,
\end{equation}
c) the condition of the
invariance of the phase coexistence line $f(a_1,a_2)=0$, which in
the canonical variables is represented by the equation $A_1=0$.

Note that similar idea of transformation of the variable reducing
the density distribution function to simpler (gaussian) form was
used in \cite{sornette/prep/2000}. In fact the existence of such a
transformation is guaranteed by the Radon-Nikodim theorem
\cite{schwartz}.

Basing on these conditions one can show (see some details in
\cite{canformkulinskii/jmolliq/2003}) that the coefficients
of the canonical form satisfy the relations:
as:
\begin{equation}
\label{a4c} A_{4} = {\frac{{\pi ^{4}}}{{\left( {\Gamma \left( {3 /
4} \right){\int\limits_{ - \infty} ^{\infty}  {dy\exp \left( { -
{\sum\limits_{n = 4}^{\infty}  {{\frac{{a_{n}}} {{n}}}y^{n}}}}
\right)}}} \right)^{4}}}}\,\,,
\end{equation}
\begin{equation}\label{a2c}
  \sqrt{\f{A_2}{A_4}}\, e^{\frac{A_2^2}{2 A_4}} K_{\frac{1}{4}}\left(\frac{A_2^2}{2
   A_4}\right) =\left.
   \int\limits^{+\infty}_{-\infty}\exp\left(-h_{l}(z)\right)dz\right|_{f(a_1,a_2)=0}\,\,.
\end{equation}
Here $K_{n}$ is the McDonald function of $n$-th kind and $\Gamma(x)$
is the Euler gamma function \cite{abramovitzstegun}.



Neglecting the variation of nonlocal part for the fluctuation part
of the Hamiltonian and after the rescaling we get the
Landau-Ginzburg (LG) form:
\begin{equation}\label{hlgcan}
\mathcal{H}_{LG}[\eta(\mathbf{r})]=\int
\left(A_{1}\eta(\mathbf{r})+\frac{A_{2}}{2}\,\eta^{2}(\mathbf{r})+
\frac{A_{4}}{4}\,\eta^{4}(\mathbf{r}) + \f{1}{2}\, (\nabla
\eta(\mathbf{r}))^2\right) dV\,\,,
\end{equation}
equivalent to the Hamiltonian of the Ising model.
Thus the possibility to transform the Hamiltonian to the canonical
form \eqref{hcan} is equivalent to the
isomorphism of the critical behavior of liquids with that for the
Ising model. Usually, for the liquids the
Hamiltonian \eqref{hreal} is obtained via grand canonical ensemble,
which means that the coefficients $a_i$ and $A_i$ are analytic functions of
the thermodynamic variables $\mu $ and $T$ \cite{yukhgol}.

Applying the standard scaling arguments (see \cite{patpokr,stanley})
to the canonical Hamiltonian \eqref{hcan} we get the fluctuation
contribution to the thermodynamic potential:
\begin{equation}\label{Phican}
  \Phi(A_1,A_2) = \Phi_0 + |A_2|^{2-\alpha}\,
  f_{s}\left(\,\f{A_1 }{|A_2|^{\beta +\gamma}} \,\right)
\end{equation}

Here $\Phi$ is the thermodynamic potential which corresponds to
the canonical coordinates $A_1, A_2$, which can be considered
as the generalized external field and the temperature conjugated
to the canonical order parameter
$\eta$. The potential $\Phi$ does not coincide with the pressure
since the order parameter $\eta$ conjugated to it is not the
density. The potential $\Phi$ is rather the function of all three
variables $P,T,\mu$ and corresponds to Eq.~\eqref{pmutinv}. Of
course, one can choose another set of variables, e.g. $P,T$. It is
in accordance with the $P,\mu,T$ invariance hypothesis mentioned
above and lead as was shown in \cite{rehrmermin/pra/1973} to the
``$t^{2\beta}$`` anomaly in the rectilinear diameter of the
density and as well as the entropy. We show below that the canonical formalism explicitly shows that both
$2\beta$ and $1-\alpha$ singularities of the rectilinear diameter of the density as the order
parameter are of the same nature. They are  generated by the asymmetry of the
initial Hamiltonian.

Note that the local transformation generates also the quasilocal
term $\eta(\nabla \eta)^2$ multiplied by the coefficient which is proportional to $\Gamma_2$.
In this case all local odd power terms as well as the term $\eta(\nabla \eta)^2$ can be canceled.
The coefficient $A_4$ is determined by the last condition. It means that the condition of the invariance
of the mean-field critical point \eqlref{a1a2} is not assumed. The coefficients $A_1$ and $A_2$ are correspondingly modified, which lead to the shift of the mean-field critical point. The appearance of such shift is connected with the inclusion of the quasilocal interaction between modes of the order parameter. It is necessary to emphasize that this shift arises before the renormalization procedure.
The detailed analysis will be the subject of separate work.

\section{Canonical form for van der Waals EOS}
Let us obtain the explicit expressions for the coefficients of the canonical form of the Hamiltonian for a
case, when 1) the density of the local Hamiltonian has the truncated $\varphi^6$-form:
\[ h_{l} (\varphi ) = {\sum\limits_{n = 1}^{6}
{{\frac{{a_{n}}} {{n}}}\varphi ^{n}}} ,
\]
2) the coefficients $a_n$ are taken from the van der Waals equation of state:
\begin{equation}\label{pvdw}
p(v,t) = \f{t}{v-1/3} - \f{9}{8v^2}\,,
\end{equation}
where $p,\,v$ and $t$ are the dimensionless pressure, the specific volume and
the temperature
\[p=P/P_c\,,\quad t = T/T_c\,,\quad v = V/V_c\]
reduced to the coordinate of the vdW critical point
(see e.g. \cite{ll5}). It can be shown that:
\[a_1 = \f{p-1+4\tau}{1+\tau}\,,\quad a_2 = 3\f{\tau}{1+\tau} \,,\quad a_3 = -3\f{\tau}{1+\tau}\,, \quad
a_4 = \f{3}{2}\f{1+9\tau}{1+\tau}\]
\[a_5 = - \f{21}{4}\f{1 + \f{81}{21} \tau}{1+\tau}\,,\quad a_6 =\f{99}{8}\,\f{1 + \f{81}{33} \tau}{1+\tau}\,. \]
Applying the procedure
described above we get:
\begin{equation}
\label{a4capprox} A_{4} \approx a^{(0)}_4 + 1.35\,\f{a^{(0)}_{6}}{\sqrt{a^{(0)}_4}}\approx 15.2
\end{equation}
where $a^{(0)}_n$ is the value of the coefficient at the critical
point.
Other coefficients are as following:
\begin{equation}\label{A1A2}
A_1 \approx\, a_1+\f{2}{3\sqrt {\pi}}\,{\frac {a_3}{\sqrt {a^{(0)}_4}}}+\f{2}{5}\,{
\frac {a^{(0)}_5}{a^{(0)}_4}}
\,,\quad
A_2\approx \, a_2\,.
\end{equation}
From \eqlref{gamma} and \eqlref{A1A2} it follows that
\begin{equation}
\label{gamma2} \Gamma _{2} \approx \f{2}{5}\,a^{(0)}_5\,\left(\,1+\f{5}{3\sqrt {\pi}}\,{\frac {a_3}{a^{(0)}_5\sqrt {a^{(0)}_4}}}\,\right)
\end{equation}
It is necessary to emphasize that these estimations are valid only in the mean field approximation.
For the accurate account of the fluctuational effects the quasilocal part of the Hamiltonian should be also taken into consideration \cite{nicoll/pra/1981}.

\section{The $2\beta$ rectilinear diameter singularity within the canonical formalism}
The coexistence curve determined by $A_1(\mu,T)=0$ in the
variables $(A_1,A_2,\eta)$ is naturally symmetrical. The canonical order parameter average value $\left\langle\,\eta
\,\right\rangle$ in accordance with Eq.~\eqref{Phican} and the
thermodynamics is:
\begin{equation}\label{etav}
  \left\langle\,\eta\,\right\rangle = -\left.
  \frac{\partial\, \Phi}{\partial\, A_1}\right|_{A_1=0} =\pm \left|
  \tilde{\tau}\right|^{\beta}\,g_s(0)\,+\ldots\,,\quad \tilde{\tau} = \left. A_2\right|_{A_1=0} = a\,t+o\left(\,
t\,\right)\,,
\end{equation}
where $g_{s}(x) = f'_{s}(x)$. The function  $f_{s}(x)$
possesses obvious symmetry $f_{s}(x) = f_{s}(-x)$, which means that
in the canonical variables the ``particle-hole`` symmetry is
restored.
The situation
changes if we return to the initial order parameter:
\[\varphi =
\eta - \f{1}{2}\,\Gamma_2\,\eta^2+\ldots \,.\]
%

Since $\eta$ is the
symmetrical order parameter, for the rectilinear diameter
$\varphi^{(d)}$ we get:
\begin{equation}\label{diamphi}
\varphi^{(d)} = -\f{1}{2}\,\Gamma_2
\left\langle\,\eta^2\,\right\rangle+ O\left(\,
\left\langle\,\eta^4\,\right\rangle\,\right)\,.
\end{equation}
From Eq.~\eqref{diamphi} and the short distance behavior of the
correlator $G(r) = \left\langle\,\eta(\m{r})\,\eta(0)
\,\right\rangle -
\left\langle\,\eta(\m{r})\,\right\rangle\,\left\langle\,\eta(0)
\,\right\rangle $ for the order parameter of the LG Hamiltonian
\eqref{hreal} which determines the critical behavior of the
fluctuation $\left\langle\,\eta^2 \,\right\rangle - \left\langle\,
\,\eta\right\rangle^2$ (see e.g. \cite{patpokr,pfeutytoulouse}):
\begin{equation}\label{eta2}
 \left\langle\,\eta^2 \,\right\rangle - \left\langle\,
\,\eta\right\rangle^2 =  G(r\to 0) = \int G(\m{k}) \, d \m{k} \sim
l_s(0)\,A_2^{1-\alpha} + \text{less singular terms}
\end{equation}
where $l_s(0)$ is some constant. Therefore, along the coexistence
curve we obtain:
\begin{equation}\label{diamphi2beta}
\varphi^{(d)} = -\frac{1}{2}\tilde{\Gamma}_{2}\left[
g_{s}^{2}(0)|\tilde{\tau}|^{2\beta
}+l_{s}(0)|\tilde{\tau}|^{1-\alpha }\right] +\text{less singular
terms}\,,
\end{equation}
where $\tilde{\Gamma}_{2}$ the value of corresponding quantity along
the coexistence curve:
\begin{equation}\label{ttau}
\tilde{\Gamma}_{2} =\left. \Gamma_2\right|_{A_1=0}\,.
\end{equation}
Thus the amplitudes $D_{i}$ are:
\begin{equation}\label{damplit}
D_{1-\alpha}  = -\f{1}{2}\,\tilde{\Gamma}_2\,l_{s}(0)\,,\quad
D_{2\beta } = -\f{1}{2}\,\tilde{\Gamma}_2\,g^2_{s}(0)\,.
\end{equation}
and either both zero or nonzeroth. Note that within such an
approach both $\tilde{\tau}^{1-\alpha}$ and
$\tilde{\tau}^{2\beta}$ anomalies contribute to the rectilinear
diameter for any parameter of state $\varphi$ chosen as the
initial order parameter, e.g. entropy, density, dielectric
permittivity etc. Thus we state that these anomalous terms are of
the same general nature and should be treated simultaneously for
the ``nondegenerate`` systems  in terms of the Catastrophe Theory.
They appear as the defect of the improper choice of the order
parameter and are generated by the asymmetrical part of the
Hamiltonian. This is different from the result of work
\cite{fishmixdiam1/pre/2003} where $t^{2\beta}$ anomaly is due to
nonlinear mixing while $t^{1-\alpha}$ anomaly is due to the linear
mixing of the fields.

It is important to note that all results discussed in this Section
do not depend on the method for the determination of the coefficient
$\Gamma_2$. In general case $\Gamma_2$ determined by the contributions
generated by both local and quasilocal terms in the initial Hamiltonian.

Note that since $l_s(0) < 0$ (see \cite{patpokr}), the amplitudes
$D_{1-\alpha}$ and $D_{2\beta}$ are of opposite signs. It is
important that their ratio is universal:
\begin{equation}\label{ratiouniv}
  \f{D_{1-\alpha}}{D_{2\beta}} = \f{l_{s}(0)}{g^2_{s}(0)}<0\,.
\end{equation}
Processing the experimental data is usually performed in terms of
the reduced temperature variable $t$, i.e.  in
Eq.~\eqref{diamphi2beta} the change $\tilde{\tau}\to t$ is made.
Therefore the ratio of corresponding amplitudes
\begin{equation}\label{Dampl}
D^{(t)}_{1-\alpha}
=D_{1-\alpha}\,a^{1-\alpha}\,,\quad D^{(t)}_{2\beta} =
D_{2\beta}\,a^{2\beta}
\end{equation}
is non universal because of Eq.~\eqref{ttau}. Therefore the ratio
is as following:
\begin{equation}\label{ratioa}
  \f{D^{(t)}_{1-\alpha}}{D^{(t)}_{2\beta}} \simeq
  a^{1-\alpha -2\beta}\,\f{l_s(0)}{g^2_s(0)}\,.
\end{equation}
where the nonuniversal factor $a$ is given by Eq.~\eqref{etav}. The
values of the amplitudes $D_{2\beta}$ and $D_{1-\alpha}$ depend on
$\tilde{\Gamma}_2$. The latter can be either positive or negative
depending on the details of the intermolecular interactions. This
may explain the fact that for molecular liquids $D_{2\beta}>0$ while
in the case of liquid metals the opposite sign takes place
\cite{fisherdiam/chemphyslet/2005,diam_schroer_weiss/statphys/2008}.
This may attributed to the
difference in the short range interaction in these systems. For the
molecular fluids the pure hard core interaction takes place while in
liquid metals the polarizability effects are essential due to
incomplete electronic shells e.g. in liquid $Hg$
\cite{anisimov,hensel/jpcondmat/1990}.
Strong polarizational effects in liquid metals and ionic fluids give
rise to screening and the formation of bound states (dimers, trimers etc.).
In such a situation one can expect that the quasilocal part of the
Hamiltonian for such systems differs very much from that of molecular
liquids because of strong dispersive interparticle interactions.

Though the functional form
$\tilde{\tau}(t)$ is non universal, it is the same for the systems,
which obey the law of corresponding states. This inference is in
correspondence with the estimates made in
\cite{fisherdiam/chemphyslet/2005} for molecular and ionic liquids
for which the density as the order parameter was used:
\begin{equation}\label{ratio}
\f{D_{1-\alpha}}{D_{2\beta}} \approx -10\,.
\end{equation}
\section{Analysis of the data for argon and water}
\begin{figure}
  \includegraphics[scale=1]{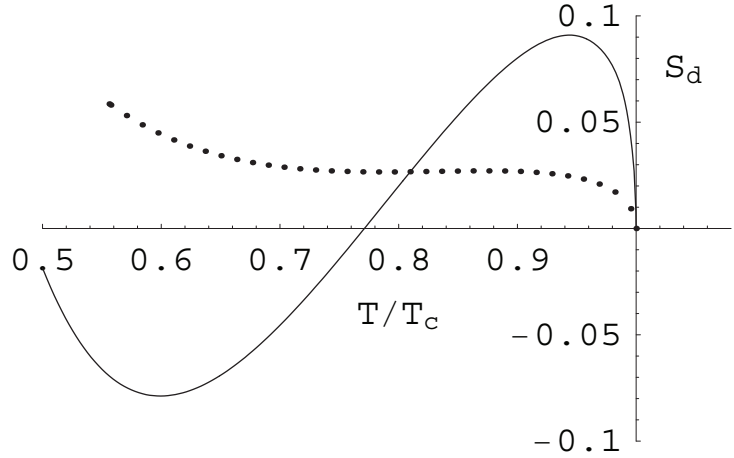}
  \caption{The rectilinear diameter of entropy %
  for water (solid) and argon (points) according %
  to data of \cite{wagnerpruss/jchrefdat/2002}.}\label{diams}
\end{figure}

To confirm the results obtained we have analyzed the experimental
data on the entropy $s$ for water and its saturated vapor
\cite{wagnerpruss/jchrefdat/2002} (see Fig.~\ref{diams}). In this
case the initial order parameter $\varphi $ is naturally defined as
$\varphi = {\frac{{s - s_{c}}} {{s_{c}}} }$. It is natural that the
entropy is more sensitive to the configuration ``hole-particle``
than the density, which is just the average of 1-particle
distribution function.

The evaporation heat $q$ and rectilinear diameter $S^{(d)}$ are
determined in the standard way: $q = T_{c} s_{c} (\varphi ^{(g)} -
\varphi ^{(l)})$ and $S_{d}(t) = {\frac{{1}}{{2}}}(S^{(g)} +
S^{(l)})$. The behavior of $\varphi ^{(d)}$ for water and argon is
presented in Fig.~\ref{diams}. In accordance with said above the
corresponding experimental data were fitted by the formulas:
\begin{equation}\label{qheat}
\frac{q}{T_{c} s_{c}} = q_{1} \vert t \vert^{\beta}  + q_{2} \vert t
\vert ^{\beta + \Delta}  + q_{3} t + o(t)
\end{equation}
and
\begin{equation}\label{diamfit}
  S^{(d)}(t) = S^{(t)}_{2\beta} \vert t \vert ^{2\beta}  +
S^{(t)}_{1-\alpha} \vert t\vert ^{1 - \alpha}  + S^{(t)}_{1}t + o(t)\,.
\end{equation}
Note, that as follows from the representation:
\begin{equation}\label{srepres}
S = c_v \, \ln T +f(n)
\end{equation}
$S_{1-\alpha} < 0 $. In contrast to the coefficients
$D_{1-\alpha}$ which is generated by the asymmetry of the
Hamiltonian, the coefficient $S_{1-\alpha}$ does not vanish in
symmetrical case, e.g. in the Ising model, and correspondingly:
\begin{equation}\label{sstruct}
 S^{(t)}_{1-\alpha} = S^{(sym)}_{1-\alpha} + S^{(asym)}_{1-\alpha}\,.
\end{equation}
It is obvious that the symmetrical part $S^{(sym)}_{1-\alpha}$ is
generated by the heat capacity singularity. The asymmetrical part
of the amplitude is of the same structure as \eqlref{Dampl} with
corresponding ratio \eqref{ratioa}.
At that we put $\alpha $ and $\Delta $ to be known: $\alpha = 0.11$
and $\Delta = 0.5$. The analysis shows that $\beta $ is the
essential function of the temperature interval $t_{\exp}$, inside
which the experimental data were taken into account.
The fitting was performed for $\beta$ and the coefficients $q_{i},
S_{i}$ in order to achieve the minimum for the standard deviation.
The value of such a minimum depends on the interval of
interpolation. Both for argon and water the optimal interval
corresponded to $0< t <t_*$ with $t_* \approx 0.2$ with close values
of coefficients though the data set for argon is smaller than that
of water. The ratio \eqref{ratioa} both for argon and water takes
close values:
\begin{equation}\label{ratiowar}
  \f{S^{(Ar)}_{1-\alpha }}{S^{(Ar)}_{2\beta}}
  \approx -3.4\,,\quad   \f{S^{(W)}_{1-\alpha }}{S^{(W)}_{2\beta}} \approx
  -3.7\,,
\end{equation}
which is in correspondence with the result \eqref{ratiouniv}. The
difference is attributed to the nonuniversal factor $a$ (see
Eq.~\eqref{ttau}) which depends on the details of the canonical
transformation. In order to distinguish between the $t^{2\beta}$ and
$t^{1-\alpha}$ terms for rectilinear diameter of the density we
subtracted the analytic background terms taking into account the
data beyond the region $t>t_{G}$. Such a subtracting allow to fix
approximately the value $t_{G}$ because the linear analytic term is
important even in fluctuational region. We fit the data in the
region $t>t^*$ with the Taylor expansion in $t$ up to $t^5$ order
and check the difference between this regular expansion and the
data. At $t^*>0.2$ no significant deviation between them was
detected.
Therefore we can conclude that the crossover takes place at $t_{G}
\approx 0.2$.

It should be noted that it is impossible to fit the data with the
the singular $t^{1-\alpha}$-term only. The $t^{2\beta}$-term is
essential and together with the analytical terms allows to fit the
data within rather broad interval $0<t< 0.2 $ correctly both for
the argon and water. Note that in approach used
\cite{chen/pra/1990,kiselevsengers/jthermophys/1993} in the
Ginzburg temperature depends on the choice of the order parameter
and may differ substantially depending on the choice of the
truncated expansion for the initial Hamiltonian.

Note that the existence of additional $t^{2\beta}$ term is also
implicitly assumed in empirical equation of state for water of
Wagner\&Pruss, which fits the data in whole coexistence region up to
the triple point with the value $\beta = 0.33$, though the term
$\sim t^{1-\alpha}$ was not taken into account
\cite{wagnerpruss/jchrefdat/2002}. From this point of view the
equations of state in broad interval, which uses the crossover
functions are preferable \cite{chen/pra/1990,aniswater/jcp/2000}.

The authors cordially thank Prof. M. A. Anisimov and Prof. H. E.
Stanley for the discussion of obtained results.

\begin{thebibliography}{29}
\expandafter\ifx\csname natexlab\endcsname\relax\def\natexlab#1{#1}\fi
\expandafter\ifx\csname bibnamefont\endcsname\relax
  \def\bibnamefont#1{#1}\fi
\expandafter\ifx\csname bibfnamefont\endcsname\relax
  \def\bibfnamefont#1{#1}\fi
\expandafter\ifx\csname citenamefont\endcsname\relax
  \def\citenamefont#1{#1}\fi
\expandafter\ifx\csname url\endcsname\relax
  \def\url#1{\texttt{#1}}\fi
\expandafter\ifx\csname urlprefix\endcsname\relax\def\urlprefix{URL }\fi
\providecommand{\bibinfo}[2]{#2}
\providecommand{\eprint}[2][]{\url{#2}}

\bibitem[{\citenamefont{Cailletet and Mathias}(1886)}]{diam0}
\bibinfo{author}{\bibfnamefont{L.}~\bibnamefont{Cailletet}} \bibnamefont{and}
  \bibinfo{author}{\bibfnamefont{E.}~\bibnamefont{Mathias}},
  \bibinfo{journal}{Scanc. Acad. Sci. Compt. Rend. Hebd., Paris}
  \textbf{\bibinfo{volume}{102}}, \bibinfo{pages}{1202} (\bibinfo{year}{1886}).

\bibitem[{\citenamefont{Sengers and Sengers}(1986)}]{diamexp/molphys/1986}
\bibinfo{author}{\bibfnamefont{J.~V.} \bibnamefont{Sengers}} \bibnamefont{and}
  \bibinfo{author}{\bibfnamefont{J.~M. H.~L.} \bibnamefont{Sengers}},
  \bibinfo{journal}{Annu. Rev. Phys. Chem.} \textbf{\bibinfo{volume}{37}},
  \bibinfo{pages}{189} (\bibinfo{year}{1986}).

\bibitem[{\citenamefont{Patashinskii and Pokrovsky}(1979)}]{patpokr}
\bibinfo{author}{\bibfnamefont{A.~Z.} \bibnamefont{Patashinskii}}
  \bibnamefont{and} \bibinfo{author}{\bibfnamefont{V.~L.}
  \bibnamefont{Pokrovsky}}, \emph{\bibinfo{title}{Fluctuation theory of
  critical phenomena}} (\bibinfo{publisher}{Pergamon},
  \bibinfo{address}{Oxford}, \bibinfo{year}{1979}).

\bibitem[{\citenamefont{Widom and Rowlinson}(1970)}]{diamrowidom/jcp/1970}
\bibinfo{author}{\bibfnamefont{B.}~\bibnamefont{Widom}} \bibnamefont{and}
  \bibinfo{author}{\bibfnamefont{J.~S.} \bibnamefont{Rowlinson}},
  \bibinfo{journal}{J. Chem. Phys.} \textbf{\bibinfo{volume}{52}},
  \bibinfo{pages}{1670} (\bibinfo{year}{1970}),
  \urlprefix\url{http://link.aip.org/link/?JCP/52/1670/1}.

\bibitem[{\citenamefont{Fisher and
  G.Orkoulas}(2000)}]{yydiamfisherorkoulas/prl/2000}
\bibinfo{author}{\bibfnamefont{M.~E.} \bibnamefont{Fisher}} \bibnamefont{and}
  \bibinfo{author}{\bibnamefont{G.Orkoulas}}, \bibinfo{journal}{Phys. Rev.
  Lett.} \textbf{\bibinfo{volume}{85}}, \bibinfo{pages}{696}
  (\bibinfo{year}{2000}).

\bibitem[{\citenamefont{Kim et~al.}(2003)\citenamefont{Kim, Fisher, and
  Orkoulas}}]{fishmixdiam1/pre/2003}
\bibinfo{author}{\bibfnamefont{Y.~C.} \bibnamefont{Kim}},
  \bibinfo{author}{\bibfnamefont{M.~E.} \bibnamefont{Fisher}},
  \bibnamefont{and} \bibinfo{author}{\bibfnamefont{G.}~\bibnamefont{Orkoulas}},
  \bibinfo{journal}{Phys. Rev. E} \textbf{\bibinfo{volume}{67}},
  \bibinfo{pages}{061506} (\bibinfo{year}{2003}).

\bibitem[{\citenamefont{Yang and Yang}(1964)}]{yang2/prl/1964}
\bibinfo{author}{\bibfnamefont{C.~N.} \bibnamefont{Yang}} \bibnamefont{and}
  \bibinfo{author}{\bibfnamefont{C.~P.} \bibnamefont{Yang}},
  \bibinfo{journal}{Phys. Rev. Lett.} \textbf{\bibinfo{volume}{13}},
  \bibinfo{pages}{303} (\bibinfo{year}{1964}).

\bibitem[{\citenamefont{Rehr and Mermin}(1973)}]{rehrmermin/pra/1973}
\bibinfo{author}{\bibfnamefont{J.}~\bibnamefont{Rehr}} \bibnamefont{and}
  \bibinfo{author}{\bibfnamefont{N.~D.} \bibnamefont{Mermin}},
  \bibinfo{journal}{Phys. Rev. A} \textbf{\bibinfo{volume}{8}},
  \bibinfo{pages}{472} (\bibinfo{year}{1973}).

\bibitem[{\citenamefont{Griffiths and
  Wheeler}(1970)}]{griffitswheeler/pra/1970}
\bibinfo{author}{\bibfnamefont{R.~B.} \bibnamefont{Griffiths}}
  \bibnamefont{and} \bibinfo{author}{\bibfnamefont{J.~C.}
  \bibnamefont{Wheeler}}, \bibinfo{journal}{Phys. Rev. A}
  \textbf{\bibinfo{volume}{2}}, \bibinfo{pages}{1047} (\bibinfo{year}{1970}).

\bibitem[{\citenamefont{Kleemeier et~al.}(1999)\citenamefont{Kleemeier,
  Wiegand, Schr{\"o}er, and Weing{\"a}rtner}}]{schroer/jchemp/1999}
\bibinfo{author}{\bibfnamefont{M.}~\bibnamefont{Kleemeier}},
  \bibinfo{author}{\bibfnamefont{S.}~\bibnamefont{Wiegand}},
  \bibinfo{author}{\bibfnamefont{W.}~\bibnamefont{Schr{\"o}er}},
  \bibnamefont{and}
  \bibinfo{author}{\bibfnamefont{H.}~\bibnamefont{Weing{\"a}rtner}},
  \bibinfo{journal}{J. Chem. Phys.} \textbf{\bibinfo{volume}{110}},
  \bibinfo{pages}{3085} (\bibinfo{year}{1999}).

\bibitem[{\citenamefont{Wagner et~al.}(2003)\citenamefont{Wagner, Stanga, and
  Schr{\"o}er}}]{schroer/pccp/2003}
\bibinfo{author}{\bibfnamefont{M.}~\bibnamefont{Wagner}},
  \bibinfo{author}{\bibfnamefont{O.}~\bibnamefont{Stanga}}, \bibnamefont{and}
  \bibinfo{author}{\bibfnamefont{W.}~\bibnamefont{Schr{\"o}er}},
  \bibinfo{journal}{Phys. Chem. Chem. Phys.} \textbf{\bibinfo{volume}{5}},
  \bibinfo{pages}{1225} (\bibinfo{year}{2003}).

\bibitem[{\citenamefont{Nicoll}(1981)}]{nicoll/pra/1981}
\bibinfo{author}{\bibfnamefont{J.~F.} \bibnamefont{Nicoll}},
  \bibinfo{journal}{Phys. Rev. A} \textbf{\bibinfo{volume}{24}},
  \bibinfo{pages}{2203} (\bibinfo{year}{1981}).

\bibitem[{\citenamefont{Nicoll and Zia}(1981)}]{nicollzia/prb/1981}
\bibinfo{author}{\bibfnamefont{J.~F.} \bibnamefont{Nicoll}} \bibnamefont{and}
  \bibinfo{author}{\bibfnamefont{R.~K.~P.} \bibnamefont{Zia}},
  \bibinfo{journal}{Phys. Rev. B} \textbf{\bibinfo{volume}{23}},
  \bibinfo{pages}{6157} (\bibinfo{year}{1981}).

\bibitem[{\citenamefont{Kulinskii}(2003)}]{canformkulinskii/jmolliq/2003}
\bibinfo{author}{\bibfnamefont{V.~L.} \bibnamefont{Kulinskii}},
  \bibinfo{journal}{J. Mol. Liq.} \textbf{\bibinfo{volume}{105}},
  \bibinfo{pages}{273} (\bibinfo{year}{2003}).

\bibitem[{\citenamefont{Sornette et~al.}(2000)\citenamefont{Sornette,
  Simonetti, and Andersen}}]{sornette/prep/2000}
\bibinfo{author}{\bibfnamefont{D.}~\bibnamefont{Sornette}},
  \bibinfo{author}{\bibfnamefont{P.}~\bibnamefont{Simonetti}},
  \bibnamefont{and} \bibinfo{author}{\bibfnamefont{J.~V.}
  \bibnamefont{Andersen}}, \bibinfo{journal}{Phys. Rep.}
  \textbf{\bibinfo{volume}{335}}, \bibinfo{pages}{19} (\bibinfo{year}{2000}).

\bibitem[{\citenamefont{Dunford and Schwartz}(1988)}]{schwartz}
\bibinfo{author}{\bibfnamefont{N.}~\bibnamefont{Dunford}} \bibnamefont{and}
  \bibinfo{author}{\bibfnamefont{J.~T.} \bibnamefont{Schwartz}},
  \emph{\bibinfo{title}{Linear Operators, General Theory}},
  vol.~\bibinfo{volume}{1} (\bibinfo{publisher}{Wiley-Interscience},
  \bibinfo{year}{1988}), \bibinfo{edition}{1st} ed., ISBN
  \bibinfo{isbn}{0471608483}.

\bibitem[{\citenamefont{Abramovitz and Stegun}(1972)}]{abramovitzstegun}
\bibinfo{author}{\bibfnamefont{M.}~\bibnamefont{Abramovitz}} \bibnamefont{and}
  \bibinfo{author}{\bibfnamefont{I.}~\bibnamefont{Stegun}},
  \emph{\bibinfo{title}{Handbook of Mathematical Functions with Formulas,
  Graphs, and Mathematical Tables}} (\bibinfo{publisher}{Dover},
  \bibinfo{address}{New York}, \bibinfo{year}{1972}).

\bibitem[{\citenamefont{Yukhnovsky and Golovko}(1980)}]{yukhgol}
\bibinfo{author}{\bibfnamefont{I.}~\bibnamefont{Yukhnovsky}} \bibnamefont{and}
  \bibinfo{author}{\bibfnamefont{M.}~\bibnamefont{Golovko}},
  \emph{\bibinfo{title}{The statistical theory of classical equilibrium
  systems}} (\bibinfo{publisher}{Naukova Dumka}, \bibinfo{address}{Kiev},
  \bibinfo{year}{1980}).

\bibitem[{\citenamefont{Stanley}(1987)}]{stanley}
\bibinfo{author}{\bibfnamefont{H.~E.} \bibnamefont{Stanley}},
  \emph{\bibinfo{title}{Introduction to Phase Transitions and Critical
  Phenomena}}, International Series of Monographs on Physics
  (\bibinfo{publisher}{Oxford University Press, USA}, \bibinfo{year}{1987}).

\bibitem[{\citenamefont{Landau and Lifshitz}(1976)}]{ll5}
\bibinfo{author}{\bibfnamefont{L.~D.} \bibnamefont{Landau}} \bibnamefont{and}
  \bibinfo{author}{\bibfnamefont{E.~M.} \bibnamefont{Lifshitz}},
  \emph{\bibinfo{title}{Statistical Physics}}, vol.~\bibinfo{volume}{5}
  (\bibinfo{publisher}{Nauka}, \bibinfo{address}{Moscow},
  \bibinfo{year}{1976}).

\bibitem[{\citenamefont{Pfeuty and Toulouse}(1977)}]{pfeutytoulouse}
\bibinfo{author}{\bibfnamefont{P.}~\bibnamefont{Pfeuty}} \bibnamefont{and}
  \bibinfo{author}{\bibfnamefont{G.}~\bibnamefont{Toulouse}},
  \emph{\bibinfo{title}{Introduction to the Renormalization Group and to
  Critical Phenomena}} (\bibinfo{publisher}{Willey\&Sons},
  \bibinfo{year}{1977}).

\bibitem[{\citenamefont{Kim and Fisher}(2005)}]{fisherdiam/chemphyslet/2005}
\bibinfo{author}{\bibfnamefont{Y.~C.} \bibnamefont{Kim}} \bibnamefont{and}
  \bibinfo{author}{\bibfnamefont{M.~E.} \bibnamefont{Fisher}},
  \bibinfo{journal}{Chem. Phys. Lett.} \textbf{\bibinfo{volume}{414}},
  \bibinfo{pages}{185} (\bibinfo{year}{2005}).

\bibitem[{\citenamefont{Weiss and
  Schr\"{o}er}(2008)}]{diam_schroer_weiss/statphys/2008}
\bibinfo{author}{\bibfnamefont{V.~C.} \bibnamefont{Weiss}} \bibnamefont{and}
  \bibinfo{author}{\bibfnamefont{W.}~\bibnamefont{Schr\"{o}er}},
  \bibinfo{journal}{Journal of Statistical Mechanics: Theory and Experiment}
  \textbf{\bibinfo{volume}{2008}}, \bibinfo{pages}{P04020 (26pp)}
  (\bibinfo{year}{2008}),
  \urlprefix\url{http://stacks.iop.org/1742-5468/2008/P04020}.

\bibitem[{\citenamefont{Anisimov}(1987)}]{anisimov}
\bibinfo{author}{\bibfnamefont{M.~A.} \bibnamefont{Anisimov}},
  \emph{\bibinfo{title}{Critical phenomena in liquids and liquid crystals}}
  (\bibinfo{publisher}{Nauka}, \bibinfo{address}{Moscow},
  \bibinfo{year}{1987}).

\bibitem[{\citenamefont{Hensel}(1990)}]{hensel/jpcondmat/1990}
\bibinfo{author}{\bibfnamefont{F.}~\bibnamefont{Hensel}}, \bibinfo{journal}{J.
  Phys. Cond. Matt. Suppl. A} \textbf{\bibinfo{volume}{2}}, \bibinfo{pages}{33}
  (\bibinfo{year}{1990}).

\bibitem[{\citenamefont{Wagner and Pru\ss}(2002)}]{wagnerpruss/jchrefdat/2002}
\bibinfo{author}{\bibfnamefont{W.}~\bibnamefont{Wagner}} \bibnamefont{and}
  \bibinfo{author}{\bibfnamefont{A.}~\bibnamefont{Pru\ss}},
  \bibinfo{journal}{J. Phys. Chem. Ref. Data} \textbf{\bibinfo{volume}{31}},
  \bibinfo{pages}{387} (\bibinfo{year}{2002}).

\bibitem[{\citenamefont{Chen et~al.}(1990)\citenamefont{Chen, Abbaci, Tang, and
  Sengers}}]{chen/pra/1990}
\bibinfo{author}{\bibfnamefont{Z.~Y.} \bibnamefont{Chen}},
  \bibinfo{author}{\bibfnamefont{A.}~\bibnamefont{Abbaci}},
  \bibinfo{author}{\bibfnamefont{S.}~\bibnamefont{Tang}}, \bibnamefont{and}
  \bibinfo{author}{\bibfnamefont{J.~V.} \bibnamefont{Sengers}},
  \bibinfo{journal}{Phys. Rev. A} \textbf{\bibinfo{volume}{42}},
  \bibinfo{pages}{4470} (\bibinfo{year}{1990}).

\bibitem[{\citenamefont{Kiselev and
  Friend}(1993)}]{kiselevsengers/jthermophys/1993}
\bibinfo{author}{\bibfnamefont{S.~B.} \bibnamefont{Kiselev}} \bibnamefont{and}
  \bibinfo{author}{\bibfnamefont{D.~G.} \bibnamefont{Friend}},
  \bibinfo{journal}{Int. Journ, Thermophys.} \textbf{\bibinfo{volume}{14}},
  \bibinfo{pages}{1} (\bibinfo{year}{1993}).

\bibitem[{\citenamefont{Wyczalkowska et~al.}(2000)\citenamefont{Wyczalkowska,
  Abdulkadirova, Anisimov, and Sengers}}]{aniswater/jcp/2000}
\bibinfo{author}{\bibfnamefont{A.~K.} \bibnamefont{Wyczalkowska}},
  \bibinfo{author}{\bibfnamefont{K.~S.} \bibnamefont{Abdulkadirova}},
  \bibinfo{author}{\bibfnamefont{M.~A.} \bibnamefont{Anisimov}},
  \bibnamefont{and} \bibinfo{author}{\bibfnamefont{J.~V.}
  \bibnamefont{Sengers}}, \bibinfo{journal}{J. Chem. Phys.}
  \textbf{\bibinfo{volume}{113}}, \bibinfo{pages}{4985} (\bibinfo{year}{2000}),
  \urlprefix\url{http://link.aip.org/link/?JCP/113/4985/1}.

\end{thebibliography}

\end{document}